\documentclass[journal,onecolumn]{IEEEtran}

\usepackage{amssymb}
\usepackage{caption2}
\usepackage{amsmath}
\usepackage{multirow}
\usepackage{amsthm}

\newtheorem{theorem}{Theorem}[section]
\newtheorem{lemma}[theorem]{Lemma}

\newtheorem{definition}[theorem]{Definition}

\newtheorem{question}[theorem]{Question}

\newcommand{\ma}{\mathcal}

\newcommand{\s}{\subseteq}

\newcommand{\fr}{\frac}
\newcommand{\lc}{\lceil}
\newcommand{\rc}{\rceil}

\begin{document}
\title{New results for traitor tracing schemes}

\author{Chong Shangguan, Jingxue Ma, Gennian Ge
\thanks{The research of G. Ge was supported by the National Natural Science Foundation of China under Grant Nos. 11431003 and 61571310.
}
\thanks{C. Shangguan is with the School of Mathematical Sciences, Zhejiang University,
Hangzhou 310027, China (e-mail: 11235061@zju.edu.cn).}
\thanks{J. Ma is with the School of Mathematical Sciences, Zhejiang University,
Hangzhou 310027, China (e-mail: majingxue@zju.edu.cn).}
\thanks{G. Ge is with the School of Mathematical Sciences, Capital Normal University,
Beijing 100048, China (e-mail: gnge@zju.edu.cn). He is also with Beijing Center for Mathematics and Information Interdisciplinary Sciences, Beijing, 100048, China.}
}
\maketitle

\begin{abstract}
    In the last two decades, several classes of codes are introduced to protect the copyrighted digital data. They have important applications in the scenarios like digital fingerprinting and broadcast encryption schemes. In this paper we will discuss three important classes of such codes, namely, frameproof codes, parent-identifying codes and traceability codes. 
    Various improvements concerning on several basic properties of these codes are presented.

    Firstly, suppose $N(t)$ is the minimal integer such that there exists a binary $t$-frameproof code of length $N$ with cardinality larger than $N$, we prove that $N(t)\ge\frac{15+\sqrt{33}}{24} (t-2)^2$, which is a great improvement of the previously known bound $N(t)\ge\binom{t+1}{2}$. Moreover, we find that the determination of $N(t)$ is closely related to a conjecture of Erd\H{o}s, Frankl and F{\"u}redi posed in the 1980's, which implies the conjectured value $N(t)=t^2+o(t^2)$. Secondly, we derive a new upper bound for parent-identifying codes, which is superior than all previously known bounds. Thirdly, we present an upper bound for 3-traceability codes, which shows that a $q$-ary 3-traceability code of length $N$ can have at most $cq^{\lc N/9\rc}$ codewords, where $c$ is a constant only related to the code length $N$. It is the first meaningful upper bound for 3-traceability codes and our result supports a conjecture of Blackburn et al. posed in 2010.
\end{abstract}


\section{Introduction}


The concept of traitor tracing scheme was introduced in 1994 by Chor, Fiat and Noar \cite{chor1994} as a method to discourage piracy. Traitor tracing schemes are useful in scenarios like digital fingerprinting and broadcast encryption schemes, where the distributed content may only be accessible to authorized users.

In \cite{ST01}, Stinson, Staddon and Wei discussed in detail four types of traitor tracing schemes, namely, frameproof codes, secure frameproof codes, parent-identifying codes and traceability codes. In this paper, we will talk about three of them except the secure frameproof codes. These codes have different traceability and are used for different purposes. For example, $t$-frameproof codes can be used to prevent a coalition of at most $t$ traitors from framing a legitimate user not in this coalition. However, they are widely
considered having no traceability for generic digital fingerprinting (it is worth mentioning that Chen and Miao \cite{chenmq} showed that frameproof codes have very good traceability for multimedia fingerprinting). Therefore, in order to trace the origin of the pirate digital content, parent-identifying codes and traceability codes are introduced, with different tracing algorithms. Applications and properties of these codes have been studied extensively, see for instance \cite{ippnogalow,ippnogaup,BL03,IPPblacknurn,BL10,Firstpaper,chenmq,chor1994,chor2000,ST01}. A major problem in this research area is to determine the upper bounds for the cardinalities of these codes. A lot of papers have been written in this aspect, see for example \cite{ippnogaup,BL03,IPPblacknurn,BL10,ST01,tranvanbound}.


Consider a code $\mathcal{C}\subseteq F^N$, where $F$ denotes an alphabet of size $q$.
Without loss of generality, we can take $F=\{0,1,\ldots,q-1\}$.
We call the code $\ma{C}$ an $(N,n,q)$ code if $|\ma{C}|=n$. Each codeword $c\in\ma{C}$ can be represented as $c=(c_1,\ldots,c_N)$, where $0\leq c_i\leq q-1$ for all $1\leq i\leq N$. Sometimes it will be more convenient if we use a matrix to describe a code. We can depict an $(N,n,q)$ code as an $N\times n$ matrix on $q$ symbols, where each column of the matrix corresponds to one of the codewords. This matrix is called the representation matrix of the code. Representation matrices of codes will be used frequently in this paper.

For any subset of codewords $D\subseteq\ma{C}$ and every $1\leq i\leq N$, we denote $desc_i(D)=\{c_i:c\in D\}$.
The set of $descendants$ of $D$ is defined as
$$desc(D)=\{x\in F^N:x_i\in desc_i(D),~1\leq i\leq N\}.$$
\noindent One can also view $desc(D)$ as $$desc(D)=desc_1(D)\times desc_2(D)\times\cdots\times desc_N(D).$$ The set $D\s\ma{C}$ is said to be a parent set of a word $x\in F^N$ if $x\in desc(D)$. We use $\ma{P}_t(x)$ to denote the collection of parent sets of $x$ such that $|D|\le t$ and $D\s\ma{C}$.

For arbitrary two vectors $x,~y\in F^n$, the Hamming distance $d(x,y)$, is defined to be the number of distinct coordinates between them:
$$d(x,y)=|\{1\le i\le N\mid x_i\neq y_i\}|.$$
\noindent Sometimes it will be more convenient to use $I(x,y)=N-d(x,y)$, which denotes the number of identical coordinates between $x$ and $y$. The minimum distance of a code $\ma{C}\s F^N$ is defined to be
$$d(\ma{C})=\min\{d(x,y)\mid x,y\in\ma{C},~x\neq y\}.$$ For a word $x\in F^N$ and a subset $D\s\ma{C}$, the group distance $d(x,D)$ is defined to be
$$d(x,D)=|\{i:1\leq i\leq N,~x_i\not\in desc_i(D)\}|.$$ \noindent Similarly, we use $I(x,D)=N-d(x,D)$ to denote the number of coordinates $i$ such that $x_i\in desc(D_i)$, $1\le i\le N$.

Now we are ready to present the definitions of the codes discussed in this paper.

\begin{definition}\label{basic}
    Suppose $\ma{C}$ is an $(N,n,q)$ code and $t\ge2$ is an integer.
    \begin{enumerate}
      \item [1)] We call $\ma{C}$ a $t$-frameproof code (or $t$-FP code for simplicity) if for all $D\subseteq\ma{C}$ with $|D|\leq t$, it holds that $$desc(D)\cap\ma{C}=D.$$
          \noindent $\ma{C}$ will be also denoted as an $FPC(N;n,q,t)$.

      \item [2)] We call $\ma{C}$ a $t$-parent-identifying code (or $t$-IPP code for simplicity) if for all $x\in F^N$, it holds that either $\ma{P}_t(x)=\emptyset$ or
          $$\cap_{D\in\ma{P}_t(x)} D\neq\emptyset.$$ \noindent $\ma{C}$ will be also denoted as an $IPP(N;n,q,t)$.

      \item [3)] We call $\ma{C}$ a $t$-traceability code (or $t$-TA code for simplicity) if for arbitrary $D\s\ma{C}$ with $|D|\le t$ and arbitrary $x\in desc(D)$, it holds that
          $$\min_{c\in D}d(x,c)<\min_{y\in\ma{C}\setminus D}d(x,y).$$ \noindent $\ma{C}$ will be also denoted as a $TA(N;n,q,t)$.

    \end{enumerate}
\end{definition}

It is well-known that $t$-TA implies $t$-IPP and $t$-IPP implies $t$-FPC. See \cite{ST01} for a more detailed description of the relations between these three codes. We have mentioned before that both the $t$-IPP codes and the $t$-TA codes can trace at least one traitor if the number of all traitors is at most $t$. Generally speaking, assume that we are given a secure code $\ma{C}$ and a coalition $D\s\ma{C}$ of at most $t$ traitors. If $x\in desc(D)$ is the pirate data, then our goal is to find some traitor $c\in D$. If $\ma{C}$ is a $t$-IPP code, we can determine $\ma{P}_t(x)$ by simply examining all small subsets (with size at most $t$) of $\ma{C}$, then the non-empty set $\cap_{E\in\ma{P}_t(x)} E$ must belong to $D$. If $\ma{C}$ is a $t$-TA code, we can find some traitor $c\in D$ by computing all distances $\{d(x,c):c\in\ma{C}\}$, then the codewords with the smallest distance must belong to $D$. To sum up, if we are given $n$ codewords with length $N$, and a coalition of at most $t$ traitors, then IPP codes and TA codes can trace at least one traitor in time $\ma{O}(N\binom{n}{t})$ and $\ma{O}(Nn)$, respectively. Note that the tracing time can be further reduced to $\ma{O}(N\log^c n)$ for some constant $c$ if a list-decoding algorithm is employed \cite{barg2004ipp,silverberg2003applications}.

A central goal in this research field is to determine the maximal cardinalities of these codes under certain fixed parameters. Let the code length $N$, the alphabet size $q$ and the strength $t$ be fixed, we use $M_{FP}(N,q,t),~M_{IPP}(N,q,t),~M_{TA}(N,q,t)$ to denote the corresponding maximal cardinalities of frameproof codes, parent-identifying codes and traceability codes. We also use $N(t)$ to denote the minimal integer $N$ such that $M_{FP}(N,2,t)>N$. Recently, the determination of $N(t)$ has received considerable attentions \cite{binaryfpc1,shangguan2014new}.

In general, there are two directions in which the bounds of these codes have been studied. The first one is to consider the maximal cardinality under small alphabet size and large code length, i.e., letting $q$ be fixed and $N$ appropriate infinity. Another direction is to consider the maximal cardinality under small code length and large alphabet size, i.e., letting $N$ be fixed and $q$ appropriate infinity. It is relatively easy to construct large codes over large alphabets, since error-correcting codes with large distance will usually satisfy the conditions of these codes \cite{ST01}. However, due to the Plotkin bound (see \cite{van2012introduction}), to construct large codes over small alphabets is much more difficult (see for example, \cite{hyperpgrahbarg,barg2004ipp,BL10,fpcDyachkov1}).

The motivation of this paper is to present improved upper bounds for these codes. We will discuss three upper bounds in total. Our first bound is a great improvement of the previously known bound for binary frameproof codes, the second one improves the known bound for parent-identifying codes and the third one is the first upper bound for 3-traceability codes. The rest of this paper is organized as follows.
The next section is devoted to frameproof codes. It is proved that $N(t)\ge\fr{15+\sqrt{33}}{24} (t-2)^2$, which improves the previously known bound $N(t)\ge\binom{t+1}{2}$ \cite{shangguan2014new}. In Section 3, we derive a new upper bound for parent-identifying codes.
And in Section 4, we present an upper bound for 3-traceability codes.
Section 5 is about some concluding remarks.

\section{Frameproof codes}

The best current bound for $M_{FP}(N,q,t)$ with small $N$ and large $q$ is due to Blackburn \cite{BL03}, who proved the following theorem:

\begin{theorem}[\cite{BL03}]\label{blackburn1}
    Let $r\in\{0,\ldots,t-1\}$ be an integer such that $r\equiv N\pmod t$. Then it holds that
    $M_{FP}(N,q,t)\leq\max\{q^{\lc N/t\rc},r(q^{\lc N/t\rc}-1)+(t-r)(q^{\lfloor N/t\rfloor}-1)\}$.
\end{theorem}

Note that the constants $r$ and $t-r$ can be reduced in many cases. For example, Corollary 9 of \cite{BL03} gives a slightly better bound with an improved constant in front of $q^{\lc N/t\rc}$, and relates this constant to a question in the theory of set systems. When $r=1$, \cite{tranvanbound} gives a clean bound $M_{FP}(N,q,t)\le q^{\lc N/t\rc}$.

We have mentioned that the determination of $M_{FP}(N,q,t)$ over small alphabets is more difficult than over the large ones. Intuitively, the most interesting and difficult case is to study the properties of binary frameproof codes. The recent papers \cite{binaryfpc2,binaryfpc1,shangguan2014new} have made some efforts on this aspect.

\begin{theorem}[\cite{binaryfpc1}] \label{binarycase}
    For all $t\geq 3$ and for all $t+1\leq N\leq 3t$, it holds that $M_{FP}(N,2,t)\le N$.
\end{theorem}

Recall that $N(t)$ is the minimal integer $N$ such that there exists an $FPC(N;n,2,t)$ with $n>N$. One can deduce from Theorem \ref{blackburn1} that $N(t)>t$ (just check the upper bound for $N\le t$). Combining this together with  Theorem \ref{binarycase} leads to the simple bound $N(t)> 3t$, which was improved to $N(t)\ge\binom{t+1}{2}$ in \cite{shangguan2014new}. Our main result on frameproof codes can be stated as the following theorem:

\begin{theorem}\label{shangguanbinary}
    For all $t\geq 3$ and for all $N<\fr{15+\sqrt{33}}{24} (t-2)^2$, it holds that $M_{FP}(N,2,t)\le N$. Or equivalently, $N(t)\ge\fr{15+\sqrt{33}}{24} (t-2)^2$.
\end{theorem}

The proof of this theorem will be postponed to the next subsection. Note that this theorem is a great improvement of both results in \cite{binaryfpc1} and \cite{shangguan2014new}. However, there is still a gap from the conjectured value $N(t)=t^2+o(t^2)$ (which will be pointed out in Section 5).

\subsection{Proof of Theorem \ref{shangguanbinary}}

Sometimes it is more convenient to use another equivalent definition for the frameproof codes.
\begin{definition}\label{fpc}
    An $(N,n,q)$ code $\ma{C}$ is a $t$-frameproof code if for every $c\in\ma{C}$ and $D\s\ma{C}$ such that $c\not\in D$ and $|D|\le t$ it holds that $c\not\in desc(D)$, which is equivalent to say that there exists some $1\le i\le N$ such that $c_i\not\in desc_i(D)$.
\end{definition}

\begin{lemma}
    The two definitions of frameproof codes are equivalent.
\end{lemma}

\begin{proof}
    On one hand, suppose $\ma{C}$ is an $(N,n,q)$ code satisfying the first statement of Definition \ref{basic}. Then given arbitrary $c\in\ma{C}$ and $D\s\ma{C}$ such that $c\not\in D$ and $|D|\le t$ it holds that $c\not\in desc(D)$, since otherwise we have $desc(D)\cap\ma{C}=D\cup\{c\}$, which violates Definition \ref{basic}.

    On the other hand, suppose $\ma{C}$ is an $(N,n,q)$ code satisfying Definition \ref{fpc}. Then given arbitrary $D\s\ma{C}$ such that $|D|\le t$ it holds that $desc(D)\cap\ma{C}=D$, since otherwise if $desc(D)\cap\ma{C}=D\cup\{c\}$ for some $c\not\in D$ and $c\in\ma{C}$, then we have $c\in desc(D)$, which violates Definition \ref{fpc}.
\end{proof}

To prove Theorem \ref{shangguanbinary}, let us introduce the definition of cover-free family. Let $X$ be a set of $N$ elements. A family $\ma{F}\s 2^{X}$ is said to be $t$-cover-free if for arbitrary distinct $t+1$ members $A_0,A_1,\ldots,A_t$ of $\ma{F}$ it holds that $A_0\nsubseteq A_1\cup A_2\cup\cdots\cup A_t$. Suppose that $|\ma{F}|=n$ and let us denote $X=\{x_1,\ldots,x_N\}$, $\ma{F}=\{A_1,\ldots,A_n\}$. $\ma{F}$ will be denoted as a $CFF(N;n,t)$. Denote by $M^*$ the representation matrix of $\ma{F}$, which is an $N\times n$ binary matrix whose rows are indexed by the elements of $X$ and whose columns are indexed by the members of $\ma{F}$, such that the entry in the $i$-th row and the $j$-th column is 1 if and only if $x_i\in A_j$. In a binary matrix, the weight of a column is simply the number of 1's contained in it.

\begin{lemma}[\cite{cffpro}]\label{cffpro}
    Let $\ma{F}$ be a $CFF(N;n,t)$ with representation matrix $M^*$. Fix an arbitrary member $A$ of $\ma{F}$ and consider the new family $\ma{F}_1$ defined by
    \begin{enumerate}
      \item [1)] $\ma{F}_1\s 2^{X\setminus A}$,
      \item [2)] $\ma{F}_1=\{B\setminus A:B\in\ma{F},~B\neq A\}$.
    \end{enumerate}
    \noindent Then $\ma{F}_1$ is a $CFF(N-|A|;n-1,t-1)$.
\end{lemma}

\begin{proof}
    The first two parameters in $CFF(N-|A|;n-1,t-1)$ are easy to verify. It suffices to prove that $\ma{F}_1$ is a $(t-1)$-cover-free family. Suppose otherwise, there are $t$ different members $B_0,B_1,\ldots,B_{t-1}$ of $\ma{F}_1$ such that $B_0\s B_1\cup\cdots \cup B_{t-1}$. For each $0\le i\le t-1$, denote $A_i$ the member of $\ma{F}$ such that $B_i=A_i\setminus A$. Then it holds that $A_0\s B_0\cup A\s (B_1\cup\cdots B_{t-1})\cup A\s A_1\cup\ldots\cup A_{t-1}\cup A$, which violates the $t$-cover-free property of $\ma{F}$.
\end{proof}

Cover-free families are closely related to binary frameproof codes. Their relations are described by the following two lemmas.

\begin{lemma}\label{cffandfpc}
    Every $CFF(N;n,t)$ is also an $FPC(N;n,2,t)$ and every $FPC(N;n,2,t)$ induces a $CFF(2N;n,t)$.
\end{lemma}

\begin{proof}
    Denote by $M$ and $M^*$ the representation matrices of an $FPC(N;n,2,t)$ and a $CFF(N;n,t)$ respectively. Given $M^*$, by the $t$-cover-free property it holds that for each column and arbitrary $t$ other columns there exists a row in which the first column is 1 and the remaining $t$ columns are all 0's. If we view the columns of $M^*$ as codewords of some binary frameproof codes, then $M^*$ surely satisfies the sufficient condition of Definition \ref{fpc}, which implies that $M^*$ also represents an $FPC(N;n,2,t)$.

    On the other hand, given $M$, then replace the 0 entry in $M$ by 10 and the 1 entry by 01. We obtain a $2N\times n$ matrix denoted by $M_1$. It suffices to verify that $M_1$ is a representation matrix of a $CFF(2N;n,t)$. Note that $M$ satisfies the $t$-frameproof property, then for each column and arbitrary $t$ other columns of $M_1$, for the corresponding columns in $M$, by the $t$-frameproof property there is a row has the configuration $10\cdots 0$ or $01\cdots 1$, which is translated to
    $$\left(\begin{array}{cccc}
        0 & 1 & \cdots & 1 \\
        1 & 0 & \cdots & 0
      \end{array}\right)~or~
      \left(\begin{array}{cccc}
        1 & 0 & \cdots & 0 \\
        0 & 1 & \cdots & 1
      \end{array}\right),
    $$
    \noindent in $M_1$. Note that the second row of the first submatrix or the first row of the second submatrix satisfies the $t$-cover-free property. If we view $M_1$ as a representation matrix of some $\ma{F}\s2^X$, where $|X|=2N$, then by the above discussions we can conclude that $\ma{F}$ is $t$-cover-free.
\end{proof}

\begin{lemma}\label{fpctightbd}
   Denote $N^*(t)$ the minimal $N$ such that there exists a $CFF(N;n,t)$ with $n>N$. And denote $N(t)$ the minimal $N$ such that there exists an $FPC(N;n,2,t)$ with $n>N$. Then for $t\ge3$, it holds that $N^*(t-2)\le N(t)\le N^*(t)$.
\end{lemma}

\begin{proof}
Denote by $M$ the representation matrix of an $FPC(N;n,2,t)$ with $N=N(t)$, then we have $n>N$ by the definition of $N(t)$. First of all, the upper bound in the inequality follows from the fact that every $CFF(N;n,t)$ is also an $FPC(N;n,2,t)$, which is shown in Lemma \ref{cffandfpc}. It remains to prove the lower bound. Replace the 0 entry in $M$ by 10 and the 1 entry by 01. We obtain a $2N\times n$ matrix with constant column weight $N$. Denote this new matrix by $M_1$. By Lemma \ref{cffandfpc}, $M_1$ is the representation matrix of a $CFF(2N;n,t)$.

By Lemma \ref{cffpro}, deleting from $M_1$ an arbitrary column and the rows containing a 1 in it leads to a new matrix $M_2$, which is the representation matrix of a $CFF(N;n-1,t-1)$.

We claim that there must exist a column in $M_2$ of weight at least two. Denote by $c$ the column deleted from $M_1$. If some column $c'\in M_2$ is of weight one, then one can verify that $c$ and $c'$ have exactly $N-1$ identical coordinates in $M$. If $M_2$ contains two columns of weight 1, then in $M$ there are two distinct columns that have exactly $N-1$ identical coordinates with $c$. Then it is not hard to prove that $c$ is contained in the descendant set of these two columns.
Therefore, $M_2$ contains at most one column of weight 1. The claim follows from the simple fact that $n-1\ge N>t\ge 3$.

Take an arbitrary column of $M_2$ with weight at least two. Delete from $M_2$ this column and the rows containing a 1 in it. Again, by Lemma \ref{cffpro}, the new matrix is the representation matrix of a $CFF(N';n-2,t-2)$ satisfying $N'\le N-2<n-2$ since we have assumed that $n>N$. Thus one can deduce that $N'\ge N^*(t-2)$ and hence the lower bound $N(t)\ge N^*(t-2)$ follows immediately.
\end{proof}

One more lemma is needed to prove Theorem \ref{shangguanbinary}.

\begin{lemma}[\cite{shangguan2015gptesting}]\label{grouptesting}
    $N^*(t)\ge \fr{15+\sqrt{33}}{24} t^2$.
\end{lemma}

\begin{proof}[\bf{Proof of Theorem \ref{shangguanbinary}}]
    Theorem \ref{shangguanbinary} is a direct consequence of Lemmas \ref{fpctightbd} and \ref{grouptesting}.
\end{proof}

\section{Parent-identifying codes}

In Section 2 we have mentioned that the upper bound of $t$-FP codes is roughly $\ma{O}(q^{\lc N/t\rc})$. However, to guarantee the traceability, $t$-IPP codes have much smaller cardinalities, which is stated as the following theorem:

\begin{theorem}[\cite{ippnogaup}]\label{ippold}
    Denote $v=\lfloor (t/2+1)^2\rfloor$, then it holds that $M_{IPP}(N,q,t)\le (v-1)q^{\lc N/(v-1)\rc}$.
\end{theorem}

A slightly worse bound was proved in \cite{IPPblacknurn} with $M_{IPP}(N,q,t)\le\fr{v(v-1)}{2}q^{\lc N/(v-1)\rc}$. In this paper, the method introduced in \cite{ippnogaup} is strengthened to prove the following theorem, which presents the best known upper bound for IPP codes.

\begin{theorem}\label{shangguanipp}
    Denote $v=\lfloor (t/2+1)^2\rfloor$ and let $0\le r\le v-2$ be a positive integer such that $N\equiv r \mod (v-1)$.
    Then it holds that $M_{IPP}(N,q,t)\le rq^{\lceil N/(v-1)\rceil}+(v-1-r)q^{\lfloor N/(v-1)\rfloor}$.
\end{theorem}

Our theorem is obviously an improvement of Theorem \ref{ippold} when $v-1\nmid N$, since the coefficient of the leading term is replaced by some constant $r<v-1$.

\subsection{Proof of Theorem \ref{shangguanipp}}

Some preparations are needed before the proof. For a vector $x\in F^N$ and a set $V\s\{1,\ldots,N\}$, a pattern of $x$ with restriction to $V$ is defined to be the ordered $|V|$-tuple written as $x|_V=(x_{i_1},\ldots,x_{i_{|V|}})$, where $i_j\in V$ for $1\le j\le |V|$ and $1\le i_1<\cdots<i_{|V|}\le N$. Let $\ma{C}$ be an $(N,n,q)$ code and $c$ be a codeword of $\ma{C}$. $c|_V$ is said to be a private pattern of $c$ if no other member of $\ma{C}$ coincides with $c$ simultaneously in all coordinates of $V$. In other words, $c|_V$ is private if $c'|_V\neq c|_V$ for any $c'\in\ma{C}\setminus\{c\}$.

Now we can prove Theorem \ref{shangguanipp} as follows.

\begin{proof}[\bf{Proof of Theorem \ref{shangguanipp}}]
    Let $\ma{C}$ be an arbitrary $IPP(N;n,q,t)$ and denote by $M$ the representation matrix of $\ma{C}$. Then $M$ is an $N\times n$ $q$-ary matrix.
    Let us partition the rows of $M$ into $v-1$ disjoint parts denoted by $V_1,\ldots,V_{v-1}$, with the property that $|V_1|=\cdots=|V_r|=\lc N/(v-1)\rc$ and $|V_{r+1}|=\cdots=|V_{v-1}|=\lfloor N/(v-1)\rfloor$. One can check that $r(\lc N/(v-1)\rc)+(v-1-r)(\lfloor N/(v-1)\rfloor)=N$ and hence $\{V_i:1\le i\le v-1\}$ is indeed a partition of the rows of $M$.

    We say a codeword $x\in\ma{C}$ is special (with respect to $\ma{C}$) if it contains some private pattern with support set $V_i$. Suppose that $|\ma{C}|\ge rq^{\lceil N/(v-1)\rceil}+(v-1-r)q^{\lfloor N/(v-1)\rfloor}+1$, our purpose is to find a subset of $\ma{C}$ which violates the $t$-identifiable parent property.

    We claim that there must exist a nonempty set $\hat{\ma{C}}\s\ma{C}$ that contains no special (with respect to to $\hat{\ma{C}}$) codewords.

    Let us delete the special codewords in $\ma{C}$ and denote the collection of the remaining codewords by $\ma{C}^{(1)}$. Second, we delete the special codewords corresponding to $\ma{C}^{(1)}$ and denote the collection of the remaining codewords by $\ma{C}^{(2)}$. Each time, whenever there is a special codeword (special among the codewords that have not been deleted yet), we delete it. We continue this procedure until we get a code $\hat{\ma{C}}$ with no special codewords in it. We claim that $\hat{\ma{C}}$ is not empty. On one hand, any pattern (particularly, with support set $V_i$) can be deleted as a private pattern of some codeword for at most one time. On the other hand, any deleted codeword (which is special in $\ma{C}^{(i)}$ for some $i\ge1$) contains at least one private pattern (corresponding to $\ma{C}^{(i)}$) with support set $V_i$. Consequently, at most $rq^{\lceil N/(v-1)\rceil}+(v-1-r)q^{\lfloor N/(v-1)\rfloor}$ special codewords can be deleted since each $V_i$ is responsible for at most $q^{|V_i|}$ distinct patterns. Taking the assumption $|\ma{C}|\ge rq^{\lceil N/(v-1)\rceil}+(v-1-r)q^{\lfloor N/(v-1)\rfloor}+1$ into account, our deletion can not delete all codewords from $\ma{C}$. Therefore, after the deletion, we are left with a nonempty set such that no codewords in it contain a private pattern (with respect to the remaining codewords) with support set $V_i$ for any $1\le i\le v-1$. Then this set satisfies the desired property mentioned in the claim. Let us take this set to be $\hat{\ma{C}}$.

    Suppose first that $t$ is even, then $v-1=(t/2+1)^2-1=t^2/4+t$. Now our aim is to pick a specified subset of $\hat{\ma{C}}$ in order to deduce the desired contradiction. We start by picking some codeword, $x_1\in\hat{\ma{C}}$. Next, we pick a codeword $x_2\in\hat{\ma{C}}$ such that $x_1|_{V_{t/2+1}}=x_2|_{V_{t/2+1}}$. Note that the property of $\hat{\ma{C}}$ guarantees the existence of such $x_2$. Denote $m_1=(t/2+1)$.

    To choose $x_3$, we consider the pattern of $x_2$ with support set $V_{2(t/2+1)}=V_{t+2}$. This pattern appears in some other codeword in $\hat{\ma{C}}$. We check whether $x_1|_{V_{t+2}}=x_2|_{V_{t+2}}$. If so, we move to the pattern with support set $V_{t+3}$ and check it. We do so until we find the first $V_i$ such that $i\ge t+2$ and $x_1|_{V_i}\neq x_2|_{V_i}$. We choose a codeword $x_3$ coinciding $x_2$ in $V_i$ and denote $m_2=i$.

    We continue this procedure. The $(k+1)$-th codeword $x_{k+1}$ is chosen as follows. Let $m_k$ be the first integer such that $m_k\ge m_{k-1}+(t/2+1)$ and $x_k|_{V_{m_k}}\neq x_i|_{V_{m_k}}$ for all $1\le i\le k-1$. Then we choose $x_{k+1}$ as the codeword coinciding $x_k$ in $V_{m_k}$. If no such $m_k$ exists, we say that $m_k$ is undefined.

    We stop when $m_k$ is undefined. Note that at most $t/2+1$ codewords can be chosen in this way since each time we skip at least $t/2+1$ patterns and there are at most $v-1=t^2/4+t=(t/2+1)t/2+t/2<(t/2+1)^2$ patterns, thus we can never pick a $(t/2+2)$-th codeword. Finally, we have picked a set $X\s\hat{\ma{C}}$ satisfying the following properties: $|X|\le t/2+1$, $x_i|_{V_{m_i}}=x_{i+1}|_{V_{m_{i}}}$ for all $1\le i\le |X|-1$.

    The descendant $s\in desc(X)$ is chosen as follows. The first $m_1=t/2+1$ patterns of $s$ (i.e. the coordinates in $V_1\cup\cdots\cup V_{m_1}$) are chosen from $x_1$, the following patterns until $V_{m_2}$ (i.e. the coordinates in $V_{m_1+1}\cup\cdots\cup V_{m_2}$) are chosen from $x_2$, and so on. The last member of $X$ contributes at most $t/2$ patterns that do not belong to the other members of $X$.

    The following observation is the core of this proof. Any $x_i\in X$ contributes at most $t/2$ patterns which do not belong to the other members of $X$. For example, fix an arbitrary $x_i\in X$. The patterns of $s\in desc(X)$ taken from $x_i$ are $V_{m_{i-1}+1},\ldots,V_{m_{i-1}+t/2},\ldots,V_{m_i}$ (for $i=1$, let $m_0=0$). By our definition of $m_i$ and $x_i$, only the first $t/2$ of the $V_i$'s, namely, $V_{m_{i-1}+1},\ldots,V_{m_{i-1}+t/2}$, could be the possible ``private" patterns of $x_i$ in $X$. Therefore, since $x_i\in\hat{\ma{C}}$, and by definition any codeword in $\hat{\ma{C}}$ contains no private pattern with support set $V_i$ for any $1\le i\le v-1$, there exists a set $Y_i=\{y_1,\ldots,y_{t/2}\}\s\hat{\ma{C}}$ with at most $t/2$ codewords such that $y_j|_{V_{m_{i-1}+j}}=x_i|_{V_{m_{i-1}+j}}$ for all $1\le j\le t/2$. So the new set $X_i$, formed by $X_i=(X\setminus\{x_i\})\cup Y_i$, can also produce the same descendant $s$, implying $s\in desc(X_i)$. Note that $|X_i|=|X|-1+|Y_i|\le t/2+1-1+t/2=t$, then it holds that $X_i\in\ma{P}_t(s)$.

    We can do the replacement similarly for all $x_k\in X$, leading to the corresponding $Y_k$'s and the newly defined $X_k$'s. Set $X_0=X$, then according to the discussion above one can see that $s\in desc(X_k)$ for all $0\le k\le |X|$. Therefore, our desired contradiction follows from the simple fact that $\cap_{0\le k\le|X|}X_k=\emptyset$ and $|X_k|\le t$, which violates the $t$-identifiable parent property.

    For odd $t$, we do exactly the same thing, only taking $m_{k+1}\ge m_k+(t+1)/2$, which gives $|X|\le(t+1)/2+1$.
 \end{proof}

\section{Traceability codes}

In the previous two sections we have described the upper bounds for FP codes and IPP codes. However, the upper bound of TA codes is much harder to determine. Despite the trivial bounds deduced from FP codes and IPP codes, the only known general upper bound for TA codes is the bound given by Blackburn et al. in \cite{BL10}:

\begin{theorem}\label{2ta}
    $M_{TA}(N,q,2)\le cq^{\lc N/4\rc}$, where $c$ is a constant only related to the code length $N$.
\end{theorem}

Unfortunately, this upper bound is also not as good as we think, since the constant $c$ is too large (larger than $N\binom{N}{\lc N/4\rc}$) compared with the constants appearing in Theorems \ref{blackburn1} and \ref{shangguanipp}. A cleaner bound $M_{TA}(4,q,2)\le 4q$ was later obtained in \cite{MR3337182}, only for 2-TA codes with length 4.

In this paper, our contribution to TA codes also concerns on the upper bound. In \cite{BL10}, the authors proposed the following question:

\begin{question}[\cite{BL10}]\label{ta}
    Let $t$ and $N$ be fixed positive integers such that $t\geq 2.$ Does there exist a constant $c$ (depending only on $t$ and $N$) such that $M_{TA}(N,q,t)\le cq^{\lceil N/t^{2}\rceil}$?
\end{question}

We answer this question positively for $t=3$. Our result can be stated as the following theorem:

\begin{theorem}\label{thm4.3}
    Let $N$ be a positive integer. Then it holds that $M_{TA}(N,q,3)\le cq^{\lc N/9\rc}$, where $c$ is a constant depending only on $N$.
\end{theorem}

\subsection{Proof of Theorem \ref{thm4.3}}
For a code $\mathcal{C}$ of length $N,$ a codeword $x \in \mathcal{C}$ and a subset $I \subseteq [N]$ of positions, where $[N]=\{1, 2, \cdots, N\},$ define
$$F_{\mathcal{C}}(x,I)=|\{y \in \mathcal{C}:y|_I=x|_I\}|.$$

\begin{lemma}\label{Lem4.2}
Let $t$ be a fixed positive integer, and let $N=9t$. Suppose that $\mathcal{C}$ is a $q$-ary $3$-traceability code of length $N$ containing three or more codewords. Then there is a set $X$ of at most $c'q^t$ codewords such that the subcode $\mathcal{C'}=\mathcal{C}\backslash X$ of $\mathcal{C}$ has $d(\mathcal{C'})\geq d(\mathcal{C})+1$, where $c'$ is a constant depending only on $N$. Note that we simply define $d(\emptyset)=\infty$.
\end{lemma}

\begin{proof}
Suppose that $d(\mathcal{C})>N-t.$ Then the Singleton bound implies that $|\mathcal{C}|\leq q^t,$ and we may take $X=\mathcal{C}$ and $\mathcal{C'}=\emptyset$ in this case. Thus we may assume that $d(\mathcal{C})\leq N-t=8t.$

Suppose that $d(\mathcal{C})\leq 2t.$ Define a subcode $\mathcal{C'}$ of $\mathcal{C}$ by removing all codewords in $\mathcal{C}$ that possess $t$ positions that are not shared with other codewords. In other words,
$$X=\{x \in \mathcal{C}: F_{\mathcal{C}}(x,I)=1 \textup{ for some $t$-subset } I \subseteq [N]\},$$ and
$$\ma{C'}=\{x\in\ma{C}:F_{\mathcal{C}}(x,I)\ge2 \textup{ for all $t$-subsets } I \subseteq [N]\}.$$
Note that $|X|=|\mathcal{C}\backslash \mathcal{C'}|\leq {N \choose t}q^t$, since there are at most $\binom{N}{t}q^t$ different $t$-tuples in a $q$-ary code of length $N$, and every codeword $x\in X$ contains at least one $t$-tuple with positions $I$ such that $F_{\ma{C}}(x,I)=1$, and such $t$-tuple belongs to exactly one $x\in X\s\ma{C}$. We only need to show that there are no distinct codewords $x,y \in \mathcal{C'}$ with $d(x,y)=d(\mathcal{C}).$ Assume, for the contrary, there are $x\neq y \in \mathcal{C'},$ such that $d(x,y)=d(\mathcal{C})\leq 2t.$ Let $I$ be a $2t$-subset of $[N]$ that contains all positions where $x$ and $y$ disagree. Then we can choose $I_1$ and $I_2$ such that $I\subseteq I_1 \cup I_2$ and $|I_1|=|I_2|=t.$ By definition of $\ma{C'}$, it holds that $F_{\mathcal{C}}(x,I_i)\geq 2$ for each $i\in\{1,2\}$, then we can also choose $y_1,y_2 \in \mathcal{C}\backslash\{x\}$ such that $x|_{I_i}=y_i|_{I_i}$ for $i=1,2$. But then $x\in desc(y,y_1,y_2),$ which contradicts the fact that $\mathcal{C}$ is a $3$-traceability code. Thus $d(\mathcal{C'})> d(\mathcal{C}),$ and so the lemma follows in this case.

Suppose that $2t<d(\mathcal{C})\leq N-t~(=8t).$ Write $d(\mathcal{C})=N-(t+\delta)$ with $0\leq \delta <6t.$ Define
$$X=\{x \in \mathcal{C}: F_{\mathcal{C}}(x,I)\le 2^{\delta+1}{N-t \choose \delta+1} \textup{ for some $t$-subset } I \subseteq [N]\},$$
and
$$\mathcal{C'}=\{x \in \mathcal{C}: F_{\mathcal{C}}(x,I)> 2^{\delta+1}{N-t \choose \delta+1} \textup{ for all $t$-subsets } I \subseteq [N]\}.$$
Note that $|X|=|\mathcal{C}\backslash \mathcal{C'}|\leq 2^{\delta+1}{N-t \choose \delta+1}{N \choose t}q^t<2^{3N}q^t$, since there are at most $\binom{N}{t}q^t$ different $t$-tuples in a $q$-ary code of length $N$, and every codeword $x\in X$ contains at least one $t$-tuple with positions $I$ such that $F_{\ma{C}}(x,I)\le 2^{\delta+1}\binom{N-t}{\delta+1}$, and such $t$-tuple belongs to at most $2^{\delta+1}\binom{N-t}{\delta+1}$ codewords $x\in X\s\ma{C}$. To prove $d(\mathcal{C'})\geq d(\mathcal{C})+1,$ it is sufficient to show that there are no distinct codewords $x,y \in \mathcal{C'}$ with $d(x,y)=d(\mathcal{C}).$ Assume, for the contrary, there are $y_0 \neq y_1 \in \mathcal{C'},$ such that $I(y_0,y_1)=t+\delta.$ Define $I_1=\{i \in [N]: y_{0,i}=y_{1,i}\}$, then we have $y_0|_{I_1}=y_1|_{I_1}$.

Take $I_2$ such that $I_1\cap I_2=\emptyset$ and $|I_2|=t$. We claim that there exists $y_2 \in \mathcal{C}$ such that $y_0|_{I_2}=y_2|_{I_2}$ and $I(y_1,y_2)\leq \delta.$ In fact, the minimum distance of $\mathcal{C}$ implies that any codeword is uniquely determined by $t+\delta+1$ of its coordinates. Once $I_2$ is fixed, it holds that

$$|\{y \in \mathcal{C}: I(y_1,y)\geq \delta+1,~y_0|_{I_2}=y|_{I_2} \}|\leq{ N-t \choose \delta+1}<F_{\mathcal{C}}(y_0,I_2).$$

\noindent The value $N-t \choose \delta+1$ means that we can choose $\delta+1$ coordinates from $[N]\backslash I_2$ such that $y_1$ and $y$ are equal, then these coordinates together with $I_2$ uniquely determine $y_1$. So, there is at least one choice for $y_2\in\ma{C}$. Now, we redefine $I_2=\{i\in [N]\backslash I_1 : y_{0,i}=y_{2,i}\},$ and write $|I_2|=t+\delta_2$ with $0\leq\delta_2\leq\delta.$ Note that $y_1$ and $y_2$ have no identical coordinates in $I_2$, since otherwise $y_0,y_1,y_2$ are identical on these coordinates and they can be added to $I_1$.

Let $D:=[N]\backslash (I_1\cup I_2).$ If $N-|I_1|-|I_2|\leq t,$ by the definition of $\mathcal{C'},$ one can choose $y_3\in \mathcal{C}\backslash\{y_0\}$ such that $y_3|_D=y_0|_D$. Thus $y_0\in desc(y_1,y_2,y_3),$ which contradicts the fact that $\mathcal{C}$ is a $3$-traceability code, and so we may assume that
$|D|>t.$ Set $J=\{i\in [N]\backslash I_1 : y_{1,i}=y_{2,i}\}.$ We have $I(y_0,\{y_1,y_2\})=|I_1|+|I_2|=2t+\delta+\delta_2$ and $I(y_1,\{y_0,y_2\})=|I_1|+|J|=t+\delta+|J|.$ We may assume $|J|\leq t+\delta_2$ since otherwise we can exchange the roles of $y_0$ and $y_1$.

Take a $t$-subset $I_3\subseteq [N]$ such that $I_3\cap (I_1\cup I_2)=\emptyset,$ and make it cover as many elements of $J$ as possible. We claim that there exists $y_3 \in \mathcal{C}$ such that $y_0|_{I_3}=y_3|_{I_3}$ and $I(y_3,\{y_1,y_2\})\leq \delta.$ As mentioned before, any codeword of $\ma{C}$ is uniquely determined by $t+\delta+1$ of its coordinates. Once $I_3$ is fixed, it holds that
$$|\{y \in \mathcal{C}: I(y,\{y_1,y_2\})\geq \delta+1, ~y_0|_{I_3}=y|_{I_3}\}|\leq 2^{\delta+1}{ N-t \choose \delta+1}<F_{\mathcal{C}}(y_0,I_3),$$
where the multiplier $2^{\delta+1}$ means that there are at most two choices for the chosen coordinates $i\in[N]\backslash I_3$, either $y|_i=y_1|_i$ or $y|_i=y_2|_i$.
So, there is at least one choice for $y_3\in\ma{C}.$ Now, we redefine $I_3=\{i\in [N]\backslash (I_1\cup I_2) : y_{0,i}=y_{3,i}\},$ and write $|I_3|=t+\delta_3$ with $0\leq\delta_3\leq\delta.$ It is not hard to show that $y_1,y_3$ and $y_2,y_3$ both have no identical coordinates on $I_3$.

\begin{figure}
\center
\begin{tabular}{ccccc}
$y_0\in \mathcal{C}'$ & $\overbrace{0000\cdots00}^{I_1}$ & $\overbrace{0000\cdots00}^{I_2}$ & $\overbrace{0000\cdots00}^{I_3}$ & $\overbrace{0000\cdots00}^{E}$\\
$y_1\in \mathcal{C}'$ & $0000\cdots00$ & $1111\cdots11$ & $1111\cdots11$ & $1111\cdots11$\\
$y_2\in \mathcal{C}$ & $\small{\ast\ast}\small{\ast\ast}\cdots\small{\ast\ast}$ & $0000\cdots00$ &$1123\cdots15$ & $\small{\ast\ast}\small{\ast\ast}\cdots\small{\ast\ast}$\\
$y_3\in \mathcal{C}$ & $\small{\ast\ast}\small{\ast\ast}\cdots\small{\ast\ast}$ & $\small{\ast\ast}\small{\ast\ast}\cdots\small{\ast\ast}$ &$0000\cdots00$ & $\small{\ast\ast}\small{\ast\ast}\cdots\small{\ast\ast}$\\
$w\in {\rm desc}(y_1, y_2, y_3)$ & $\underbrace{0000\cdots00}_{|I_1|=t+\delta}  $ & $\underbrace{0000\cdots00}_{|I_2|=t+\delta_2}$ & $\underbrace{0000\cdots00}_{|I_3|=t+\delta_3}$ & $\underbrace{1111\cdots11}_{|E|=6t-\delta-\delta_2-\delta_3}$
\end{tabular}
\caption{When $4t\leqslant \delta < 6t$}
\end{figure}

For $\{i,j,k\}=\{1,2,3\}$, we denote $I_{i,j,k}:=\{u\in I_i : y_{j,u}=y_{k,u}\},$ then one can deduce $|I_{1,0,2}|\leq I(y_1,y_2)\leq \delta,$ and $|I_{1,0,3}|+|I_{2,0,3}|\leq I(y_3,\{y_1,y_2\})\leq \delta.$ Write $E:=[N]\backslash (I_1\cup I_2 \cup I_3)$. It is easy to see $|E|=6t-\delta-\delta_2-\delta_3$ and $|E|>0$, since otherwise $y_0\in desc(y_1,y_2,y_3)$, which contradicts the definition of 3-traceability. In the following we will consider two cases where $4t\leq\delta<6t$ and $0\leq\delta<4t.$

{\bf Case 1: $4t\leq\delta<6t$}

In this case, we take a word $w\in desc(y_1, y_2, y_3)$ with $w|_E=y_1|_E$ and $w|_{I_j}=y_j|_{I_j}$, where $j=1,2,3.$ Such choice for $w$ is well-defined since $E\cup(\cup_{i=1}^3 I_j)=[N]$ and they are all pairwise disjoint. See Figure 1 for an illustration of our notation. It is easy to compute the following inequalities:

\begin{equation*}\label{eqs1}
\left\{
\begin{array}{rcl}
I(y_0,w)&=&|I_1|+|I_2|+|I_3|=3t+\delta+\delta_2+\delta_3,\\
I(y_1,w)&=&|I_1|+|E|=(t+\delta)+(6t-\delta-\delta_2-\delta_3)=7t-\delta_2-\delta_3\\ &\leq& 7t \leq3t+\delta \leq I(y_0,w),\\
I(y_2,w)&\leq&|I_{1,0,2}|+|I_2|+|E|\le \delta+(t+\delta_2)+(6t-\delta-\delta_2-\delta_3)=7t-\delta_3\\&\leq& 7t \leq3t+\delta\leq I(y_0,w),\\
I(y_3,w)&\leq&|I_{1,0,3}|+|I_{2,0,3}|+|I_3|+|E|\le\delta+(t+\delta_3)+(6t-\delta-\delta_2-\delta_3)= 7t-\delta_2\\&\leq& 7t \leq3t+\delta\leq I(y_0,w).
\end{array}\right.
\end{equation*}

Since $y_0\not\in\{y_1,y_2,y_3\}$, this contradicts the 3-traceability property of $\ma{C}$, as required.

{\bf Case 2: $0\leq\delta<4t$}

In this case, as the above, we take a word $w\in desc(y_1, y_2, y_3)$ with $w|_{I_j}=y_j|_{I_j}$, where $j=1,2,3.$ However, we should be more careful about the choice of $w|_E$.

For $1\le i<j\le3$, define $J_{i,j}:=\{t\in E : y_{i,t}=y_{j,t}\}$ and $J_{1,2,3}:=J_{1,2}\bigcap J_{1,3}\bigcap J_{2,3}.$ We have $|J_{1,2,3}|\leq |J_{1,2}|\leq \textup{max} \{|J|-t,0\} \leq\delta_2$, since $J_{1,2,3}\s J_{1,2}\s J\setminus I_3$ and we have chosen $y_3$ to cover as many elements of $J$ as possible. Taking into account the fact that $|J_{1,3}\backslash J_{1,2,3}|+|J_{2,3}\backslash J_{1,2,3}|\leq I(y_3,\{y_1,y_2\})\leq \delta <4t,$ we consider the following two subcases separately.

{\bf Subcase 2.1}: $|J_{1,3}\backslash J_{1,2,3}|\leq 2t+\delta_3$ and $|J_{2,3}\backslash J_{1,2,3}|\leq 2t+\delta_3$

\begin{figure}
\center
\begin{tabular}{ccccc}
$y_0\in \mathcal{C}'$ & $\overbrace{0000\cdots00}^{J_{1,2}}$ & $\overbrace{0000\cdots00}^{J_{1,3}\backslash J_{1,2,3}}$ & $\overbrace{0000\cdots00}^{J_{2,3}\backslash J_{1,2,3}}$ & $\overbrace{0000\cdots00}^{H}$\\
$y_1\in \mathcal{C}'$ & $1111\cdots11$ & $1111\cdots11$ & $1111\cdots11$ & $1111\cdots11$\\
$y_2\in \mathcal{C}$ & $1111\cdots11$ & $2222\cdots22$ &$2222\cdots22$ & $2222\cdots22$\\
$y_3\in \mathcal{C}$ & $1231\cdots23$ & $1111\cdots11$ &$2222\cdots22$ & $3333\cdots33$\\
$w\in {\rm desc}(y_1, y_2, y_3)$ & $\underbrace{1111\cdots11}_{|J_{1,2}|\leq\delta_2}  $ & $\underbrace{2222\cdots22}_{|J_{1,3}\backslash J_{1,2,3}|\leq 2t+\delta_3}$ & $\underbrace{1111\cdots11}_{|J_{2,3}\backslash J_{1,2,3}|\leq 2t+\delta_3}$ & $\underbrace{\small{\ast\ast}\small{\ast\ast}\cdots\small{\ast\ast}}_{|H|=|E|-|J_{1,2}\bigcup J_{1,3}\bigcup J_{2,3}|}$
\end{tabular}
\caption{When $0\leqslant \delta < 4t$, $|J_{1,3}\backslash J_{1,2,3}|\leq 2t+\delta_3$ and $|J_{2,3}\backslash J_{1,2,3}|\leq 2t+\delta_3$}
\end{figure}

We give steps as follows to define $w_i$ when $i\in E.$ See Figure 2 for an illustration of our notation.

\begin{enumerate}
 \item  Take $w_i=y_{1,i}$, when $i\in J_{1,2}\bigcup J_{2,3},$
 \item  Take $w_i=y_{2,i},$ when $i \in J_{1,3}\backslash J_{1,2,3},$
 \item  For the remaining coordinates, i.e., these coordinates in $H:=E\backslash(J_{1,2}\bigcup J_{1,3}\bigcup J_{2,3})$. Note that each pair of $y_1|_H$, $y_2|_H$ and $y_3|_H$ has no identical coordinates, we may partition $H$ into three disjoint parts, $H_1,H_2,H_3$, satisfying the property that $|H_1|\le 2t+\delta_3-|J_{2,3}\backslash J_{1,2,3}|,$ $|H_2|\le 2t+\delta_3-|J_{1,3}\backslash J_{1,2,3}|,$ and $|H_3|=|H|-|H_1|-|H_2|\leq 2t.$ To see that such partition does exist, note that the first two inequalities are valid by our assumption on the sizes of $J_{1,3}\backslash J_{1,2,3}$ and $J_{2,3}\backslash J_{1,2,3}$, and the third inequality holds since $|H_3|\le |E|-(|H_1|+|J_{2,3}\backslash J_{1,2,3}|)-(|H_2|+|J_{1,3}\backslash J_{1,2,3}|)$, $|E|=6t-\delta-\delta_2-\delta_3$, and one can choose $|H_1|+|J_{2,3}\backslash J_{1,2,3}|$ and $|H_2|+|J_{1,3}\backslash J_{1,2,3}|$ as large as $2t+\delta_3$. For the undefined coordinates of $w$, we take $w|_{H_j}=y_j|_{H_j}$, where $j=1,2,3$. One can verify that such choice for $w$ is well-defined.
\end{enumerate}

Now, we compute the values $I(w,y_{j})$ for $j\in \{0,1,2,3\}.$ Recall that $|J_{1,2}|\le\delta_2$, $|I_{1,0,2}|+|J_{1,2}|\le I(y_1,y_2)\le\delta$ and $|I_{1,0,3}|+|I_{2,0,3}|+|J_{1,2,3}|\le I(y_3,\{y_1,y_2\})\le\delta$, then we have
\begin{equation*}\label{eqs1}
\left\{
\begin{array}{rcl}
I(y_0,w)&=&|I_1|+|I_2|+|I_3|=3t+\delta+\delta_2+\delta_3,\\
I(y_1,w)&=&|I_1|+|J_{1,2}|+|J_{2,3}\backslash J_{1,2,3}|+|H_1|\le(t+\delta)+\delta_2+(2t+\delta_3)\\
&=& 3t+\delta+\delta_2+\delta_3 = I(y_0,w),\\
I(y_2,w)&=&|I_{1,0,2}|+|I_2|+|J_{1,2}|+|J_{1,3}\backslash J_{1,2,3}|+|H_2|\leq\delta+(t+\delta_2)+(2t+\delta_3)\\
&=& 3t+\delta+\delta_2+\delta_3 = I(y_0,w),\\
I(y_3,w)&=&|I_{1,0,3}|+|I_{2,0,3}|+|I_3|+|J_{1,2,3}|+|H_3|\leq\delta+(t+\delta_3)+2t\\
&=& 3t+\delta+\delta_3\leq I(y_0,w).
\end{array}\right.
\end{equation*}

Since $y_0\not\in\{y_1,y_2,y_3\}$, this contradicts the 3-traceability property of $\ma{C}$, as required.

{\bf Subcase 2.2:} Without loss of generality, we can assume $|J_{2,3}\backslash J_{1,2,3}|>2t+\delta_3.$ Thus we define $J_{2,3}'$ such that $J_{2,3}'\subseteq J_{2,3}\backslash J_{1,2,3}$ and $|J_{2,3}'|=2t+\delta_3.$

\begin{figure}
\center
\begin{tabular}{ccc}
$y_0\in \mathcal{C}'$ & $\overbrace{0000\cdots00}^{J_{2,3}'}$ & $\overbrace{0000\cdots00}^{E\backslash J_{2,3}'}$\\
$y_1\in \mathcal{C}'$ & $1111\cdots11$ & $\small{\ast\ast}\small{\ast\ast}\cdots\small{\ast\ast}$\\
$y_2\in \mathcal{C}$ & $2222\cdots22$ & $2222\cdots22$\\
$y_3\in \mathcal{C}$ & $2222\cdots22$ & $\small{\ast\ast}\small{\ast\ast}\cdots\small{\ast\ast}$\\
$w\in {\rm desc}(y_1, y_2, y_3)$ & $\underbrace{1111\cdots11}_{|J_{2,3}'|=2t+\delta_3}$ & $\underbrace{2222\cdots22}_{|E\backslash J_{2,3}'|<2t}$
\end{tabular}
\caption{When $0\leqslant \delta<4t$ and $|J_{2,3}\backslash J_{1,2,3}|>2t+\delta_3$}
\end{figure}

Now, we give steps as follows to define $w_i$ when $i\in E.$ See Figure 3 for an illustration of our notation.

\begin{enumerate}
 \item  Take $w_i=y_{1,i}$, when $i\in J_{2,3}',$
 \item  Take $w_i=y_{2,i},$ when $i \in E\backslash J_{2,3}'.$
\end{enumerate}

Now, we compute the values $I(w,y_{j})$ for $j\in \{0,1,2,3\}.$ Since $2t+\delta_3<|J_{2,3}\backslash J_{1,2,3}|\le I(y_3,\{y_1,y_2\})\leq \delta <4t$ and $|E\backslash J_{2,3}'|=\textup{max}\{4t-\delta-\delta_2-2\delta_3,0\}< 2t,$ we have
\begin{equation*}\label{eqs1}
\left\{
\begin{array}{rcl}
I(y_0,w)&=&|I_1|+|I_2|+|I_3|=3t+\delta+\delta_2+\delta_3,\\
I(y_1,w)&\le&|I_1|+|J_{1,2}|+|J_{2,3}'|\leq (t+\delta)+\delta_2+(2t+\delta_3)=3t+\delta+\delta_2+\delta_3\\
 &=& I(y_0,w),\\
I(y_2,w)&=&|I_{1,0,2}|+|I_2|+|E\backslash J_{2,3}'|<\delta+(t+\delta_2)+2t=3t+\delta+\delta_2\\&\leq& I(y_0,w),\\
I(y_3,w)&\leq&|I_{1,0,3}|+|I_{2,0,3}|+|I_3|+|E\backslash J_{2,3}'|<\delta+(t+\delta_3)+2t=3t+\delta+\delta_3\\
&\leq& I(y_0,w).
\end{array}\right.
\end{equation*}

Since $y_0\not\in\{y_1,y_2,y_3\}$, this contradicts the 3-traceability property of $\ma{C}$, as required.
\end{proof}

\begin{proof}[\bf{Proof of Theorem \ref{thm4.3}}]
Write $N=9t-r,$ where $t\in \mathbb{Z}$ and $0\leq r\leq8.$ By concatenating all codewords with the word $0^{r}$, we may regard $\mathcal{C}$ as a traceability code of length $9t$. So we may assume that $N$ is divisible by $9$. Let $d=d(\mathcal{C})$. By applying Lemma \ref{Lem4.2} at most $N-d$ times, we obtain a code $\mathcal{C'}$ with minimal distance $N$, which has at most $q$ codewords. We have removed at most $(N-d)c'q^t$ codewords to obtain $\mathcal{C'}$, and so $|C|\leq (N-d)c'q^t +q\leq cq^t$ where we define $c=Nc'.$ So the theorem follows.
\end{proof}

\section{Concluding remarks}
In this paper, we present several new upper bounds for different traceability schemes. There are two problems remaining open.

The first one is to determine the exact (or asymptotic) value of $N(t)$. It was conjectured by Erd\H{o}s, Frankl and F{\"u}redi \cite{CFF} that $N^*(t)=t^2+o(t^2)$. If this conjecture is true, then it follows that $N(t)=t^2+o(t^2)$ by Lemma \ref{fpctightbd}. There is still a gap between the best known value and the conjectured one.

The second problem is to answer Question \ref{ta}. An interesting property that both FP codes and IPP codes satisfy is the composition law, which states that $M_{FP}(aN,q,t)<M_{FP}(N,q^a,t)$ and $M_{IPP}(aN,q,t)<M_{IPP}(N,q^a,t)$ hold for every positive integer $a$. This property says that an $FPC(N;n,q^a,t)$ (resp. an $IPP(N;n,q^a,t)$) exists if only an $FPC(aN;n,q,t)$ (resp. an $IPP(aN;n,q,t)$) exists. This composition law can be proved directly by splitting a codeword of length $aN$ into $N$ blocks of $a$ coordinates each and then viewing this codeword as a vector of length $N$ over an alphabet of size $q^a$. Unfortunately, because of the minimum distance condition required in its definition, TA codes do not seem to satisfy such a law. This may be one reason why the upper bound of TA codes is hard to estimate. It seems that our method in proving Theorem \ref{thm4.3} can be further generalized, with a more complicated discussion.

\bibliographystyle{plain}
\bibliography{recentprogress}

\end{document}